# Superposition of waves for modeling COVID-19 epidemic in the world and in the countries with the maximum number of infected people in the first half of 2020

**E.M. Koltsova[1, a], E.S. Kurkina[1, 2, b], A.M. Vasetsky[1, c]**

[1] Mendeleev University of Chemical Technology of Russia, Moscow, the Russian Federation
[2] Lomonosow Moscow State University, Moscow, the Russian Federation
[a] E-mail: kolts@muctr.ru
[b] E-mail: e.kurkina@rambler.ru
[c] E-mail: amvas@muctr.ru

**Abstract.** On the base of logic discrete equations system mathematical modeling of COVID-19 epidemic spread was carried out in the whole world and in the countries with the largest number of infected people such as the USA, Brasil, Russia and India in the first half of 2020. It was shown that for the countries with strong restrictive measures the spread of COVID-19 fit on a single wave with small capacity as for a number of countries with violation of restrictive measures the spread of the epidemic fit on a waves superposition. For countries with large population mixing (Brazil, India), the spread of the coronavirus epidemic today also fits into a single wave, but with a huge capacity value (for Brazil - 80 million people, for India - 40 million people). We estimated that the epidemic spread in the world today fits into 5 waves. The first two waves are caused by the epidemic spread in China (the first - in Wuhan), the third - by the epidemic spread in European countries, the fourth mainly by the epidemic spread in Russia and the USA, the fifth wave is mainly caused by the epidemic spread in Latin America and South Asia. It was the fifth wave that led to the spread of the coronavirus epidemic COVID-19 entering a new phase, with an increase in the number of infected more than 100 thousand inhabitants. For all the studied countries and the world, for each of the superposition waves, the wave capacities and growth indicators were calculated. The local peaks of the waves and their ending times are determined. It was the fifth wave that led to the fact that COVID-19 spread is entering a new phase, with the increase in the number of infected people being more than 100 thousand inhabitants. For all the countries being examined and for the whole world, for each of the superposition waves we calculated the waves capacities and index of infected people growth. The local peaks of the waves and their ending times are determined.

**Acknowledgements:** The authors are grateful to the postgraduate student A.S. Shaneva of D. Mendeleev University for help in editing the article.

### 1. Introduction

The major part of the papers devoted to modeling the COVID-19 epidemic spread is based on the mathematical model named SIR-model. SIR – model (*Susceptible-Infectious-Recovered* model) is considered to be more realistic. According to SIR – model the total population is divided into three parts: the proportion of the people susceptible to the disease - *y (Susceptible)*, infected - *x (Infectious)* and recovered - *z (Recovered)*. The presence of recovering people in the model reduces the number of infected people who are able to infect others.

The equations of the SIR model have the form:

$$\frac{dy}{dt} = -\lambda y \frac{x}{N},$$
$$\frac{dx}{dt} = \lambda y \frac{x}{N} - \mu x, \qquad (1)$$
$$\frac{dz}{dt} = \mu x,$$

where *y* – the part of people susceptible to the disease, *x* – the number of infected inhabitants, *z* – the number of recovered people, $\lambda$ – the index of infected growth, $\mu$ – the index of recovered growth.

This SIR-model was used earlier in the papers [1, 2] for modeling infection spread among people and animals. In this model epidemic spread is determined by the ratio of the parameters $\lambda/\mu$. If this ratio is less than 1, the epidemic is fading; if it is more than 1, the epidemic is spreading. For modeling of COVID-19 epidemic spread SIR- model was used in the [3-5].

There the SIR model development - the SEiR model. In this model, the group of infected people is divided into two subgroups. Group 1 is at the stage of the incubation period and is not contagious; Group 2 is made up of people who are contagious. Thus, the transition S → E → I → R occurs. SEiR-model as a software package is presented on an electronic resource [6]. Many authors on mathematical modeling of COVID-19 epidemic spread used the extended SEiR-model (with a number of additional states of infected and recovered groups), implemented by researcher R. Neira team and presented in the public domain on an electronic resource [7].

The SEiR-model requires the determination of many parameters. The task of model parameters searching with the use of actual data is the inverse problem and incorrect, i.e. having several solutions. And if the task is to determine only the number of infected people and the increase in the number of infected, it is better to use the discrete logistic equation (for the application of which it is necessary to determine only two parameters - the index of growth in the number of infected people and the capacity of the system).

The papers [8, 9] present the results of discrete logistic equation use to simulate the epidemic spread in Moscow (at the initial stages of the epidemic), in a number of European, Asian countries, and Israel. Modeling the epidemic spread not only at the initial stage, we noticed that the use of the only one discrete logistic equation is well applicable before the peak and on the peak of epidemic spread.

And in those countries where, after the peak of the increase in the number of infected people, the so-called "plateau" occurs and there is no drop in the increase in the number of infected people (as new sites of infection appear), we have to increase the capacity of the system all the time. So the model lost its predictive power, and mismatching between the calculated and actual data increased significantly. The parameters were adjusted to the right side (after the peak) of the curve of the increase of the infected people and we obtained the mismatching from the left side of the curve involved. And only the use of not a single but a system of discrete logistic equations led to the fact that the presence of waves superposition of reduced significantly the mismatching between the calculated and actual data both for the curve of infected people number and for the curve of the increase in the number of infected people.

Using the waves superposition approach we have achieved the smallest mismatching between actual and calculated data on the examples of modeling the epidemic spread in such countries as Spain and Italy (which are already almost completing the epidemic), in this work we consider the epidemic spread in the countries with the highest increase in the number of infected: the USA, Brazil, Russia, India as well as the epidemic spread all over the world.

## 2. Mathematical model

For modeling with the use of superposition we have the system of discrete logistic equations:

$$y_{n+1}^{(i)} = \lambda^{(i)} y_n^{(i)} (1 - y_n^{(i)}/N^{(i)}) \qquad (2)$$

$y_n^{(i)}$ – the number of inhabitants in the *n*-th moment of time who has infected during the *i*-th wave of the epidemic spread, $N^{(i)}$ is the system capacity that characterizes the maximum value for the number of people who can potentially be infected under the *i*-th wave of the epidemic spread. This parameter *N* depends on the whole number of factors such as the country openness to the virus infection (the influx of people from the epidemic centers), crowding, population density, the presence of megacities, population mobility and mixing, the presence of transport hubs, the discipline of the population, disease resistance, etc.

The model describes the growth in accordance with the logistic function: the number of patients $y_n^{(i)}$ in the *i*-th wave of epidemic spread grows dramatically until it is rather small ($y_n^{(i)} \ll N^{(i)}$), after that the growth rate is slower and slower and the number of patients asymptotically tends to a stationary value $\bar{y}^{(i)}$. Let's make a change in variables and bring the system (2) to the form

$$x_{n+1}^{(i)} = \alpha^{(i)} x_n^{(i)} \left(1 - x_n^{(i)}\right), x_0^{(i)} = y_0^{(i)}/N^{(i)}, \tag{3}$$

where variables $x_n^{(i)}$ and parameters $\alpha^{(i)}$ are dimensionless. When values $0 < \alpha^{(i)} \leq 1$ regardless of the initial value of $x_0^{(i)}$, the number of infected people during the $i$-th wave will tend to zero. When the values $1 < \alpha^{(i)} \leq 3$ regardless of the initial state, dimensionless number of the population infected during the $i$-th wave of epidemic spread will tend to the stationary steady state.

$$\bar{x}^{(i)} = \frac{\alpha^{(i)} - 1}{\alpha^{(i)}} \tag{4}$$

Therefore during the time the number of people infected in the $i$-th wave of epidemic spread (at the time of its completion) will be equal

$$\bar{y}^{(i)} = \frac{(\alpha^{(i)} - 1)}{\alpha^{(i)}} \cdot N^{(i)} \tag{5}$$

Relations (4) and (5) are true while maintaining a constant index for the growth of the population $\alpha^{(i)}$ during all the time of the $i$-th epidemic wave spread. For small numbers of infected people (at the initial stages of the i-th wave), we have the equality

$$\frac{x_{n+1}^{(i)}}{x_n^{(i)}} = \alpha^{(i)} \tag{6}$$

or

$$\frac{y_{n+1}^{(i)}}{y_n^{(i)}} = \alpha^{(i)} \tag{7}$$

From the relations (6) and (7) we can see that the capacity of the wave $N^{(i)}$ at the initial period of wave spread does not affect on the calculation of the number of infected people.

We will write the relation for determining the number of infected people at the local peak for the increase in infected people in the i-th wave of the epidemic spread

$$x_{max}^{(i)} = \frac{\alpha^{(i)} - 1}{2\alpha^{(i)}} \tag{8}$$

and
$$y_{max}^{(i)} = \left(\frac{\alpha^{(i)}-1}{2\alpha^{(i)}}\right) \cdot N^{(i)} \qquad (9)$$

We can see from the expression (9) that the system capacity $N^{(i)}$ can be defined in the vicinity of a local peak of the $i$-th wave of epidemic spread.

The system capacity will be equal

$$N = \sum_{i=1}^{M} N^{(i)}, \qquad (10)$$

where $M$ – the number of epidemic waves for the particular country. The number of all the infected people is determined according to the relation

$$\bar{y} = \sum_{i=1}^{M} \bar{x}^{(i)} N^{(i)} \qquad (11)$$

We have developed the algorithm for searching the parameters $(\alpha^{(i)}, N^{(i)})$ on the base of actual data on the number of infected people as well as on the daily increase in the number of infected that provides the minimum mismatching between calculated and actual data on the number of infected and on the increase of the number of infected.

## 3. Calculation results

Figure 1 shows the results on daily increase of COVID-19 infected population in China.

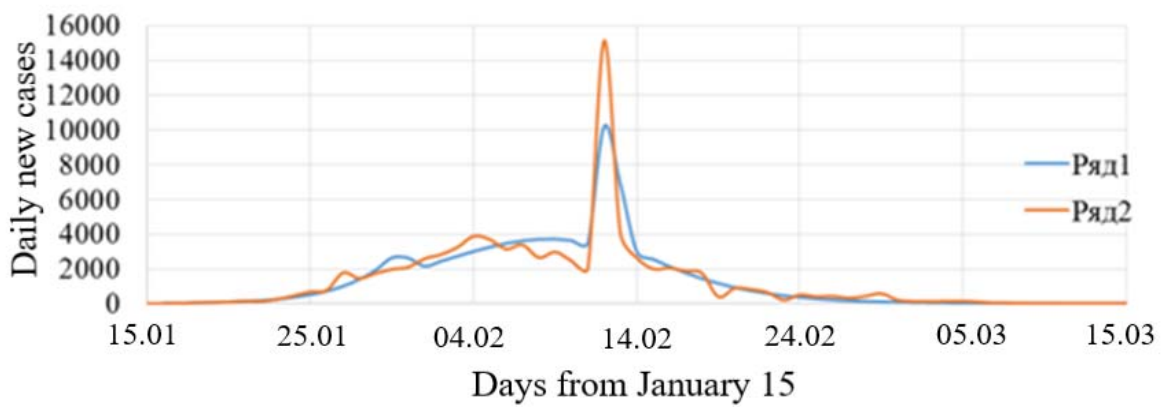

Fig. 1. Results on daily increase of infected in China

Fig. 2 – 3 present the results of calculation in daily increase of infected people at the initial stage (fig.2) and at the final stage (fig.3) for epidemic development in Moscow.

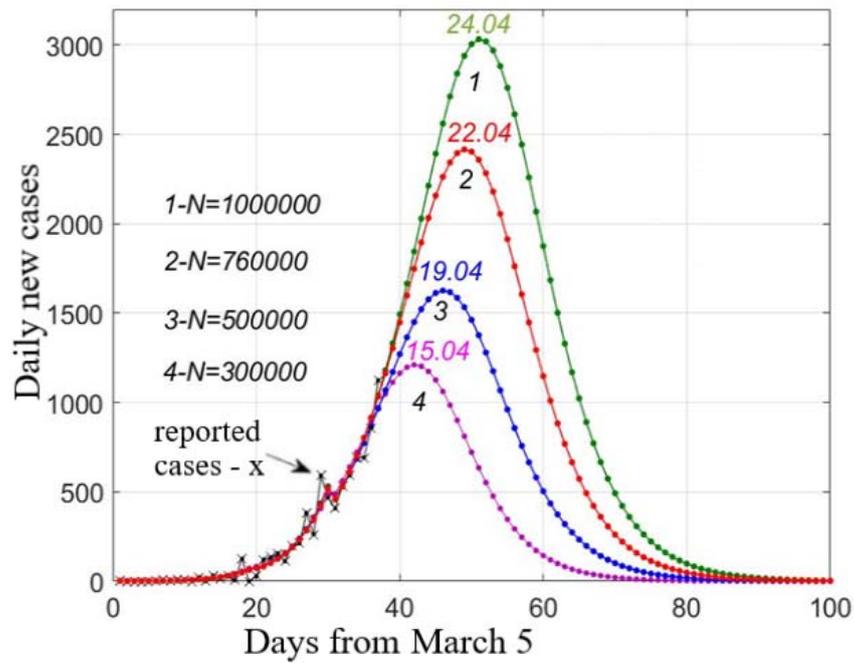

Fig. 2. The initial stage of the epidemic development in Moscow according to the 4-th scenarios. The results of the daily increase in the number of infected, where ◊ – actual data [12]

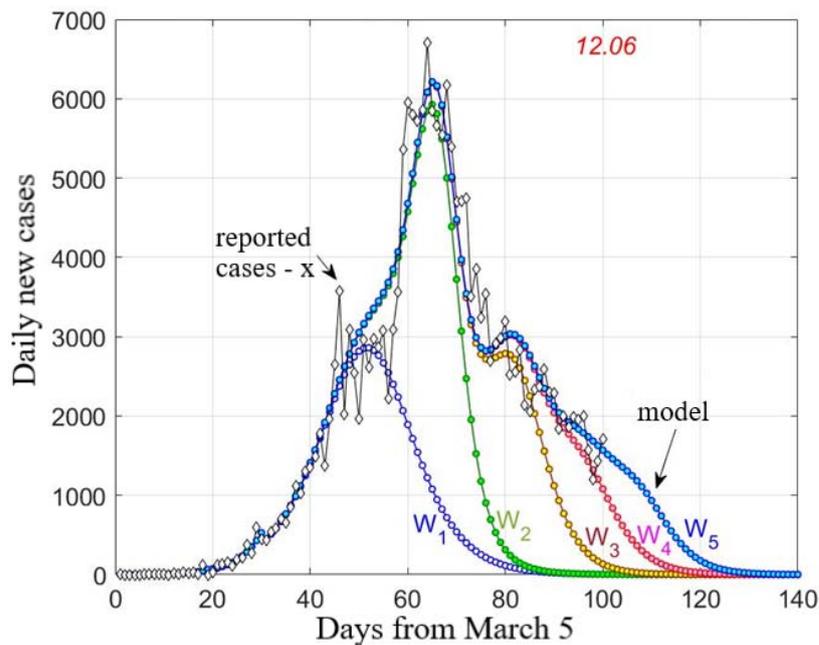

Fig. 3. All the stages of coronovirus epidemic development in Moscow. The results of the daily increase in the number of infected. Where ◊ – actual data [12]; $W_1$ – the results of the 1-st wave; $W_2$ – the results of superposition of the 1-st and 2-nd waves; $W_3$ – the result of superposition of the 1, 2 and 3-d waves; $W_4$ – the results of superposition of the 1, 2, 3, 4-th waves; $W_5$ – the results of superposition of the 1, 2, 3, 4 and 5-th waves

Fig. 4 – 7 show the calculation results on the daily increase in the number of COVID-19 infected people in such countries as: Russia, India, Brasil the USA.

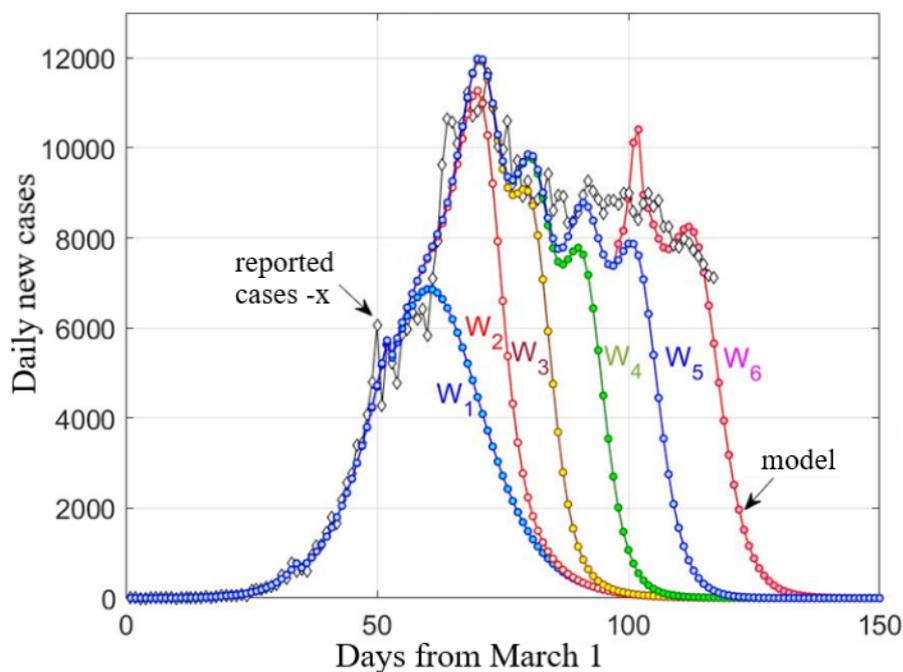

Fig. 4. All the stages of coronovirus epidemic development in Russia. The results of the daily increase in the number of infected. Where ◊ – actual data [12]; $W_1$ – the results of the 1-st wave; $W_2$ – the results of superposition of the 1-st and 2-nd waves; $W_3$ – the result of superposition of the 1, 2 and 3-d waves; $W_4$ – the results of superposition of the 1, 2, 3, 4-th waves; $W_5$ – the results of superposition of the 1, 2, 3, 4 and 5-th waves; $W_6$ – the results of the superposition of the 1, 2, 3, 4, 5 and the 6-th waves

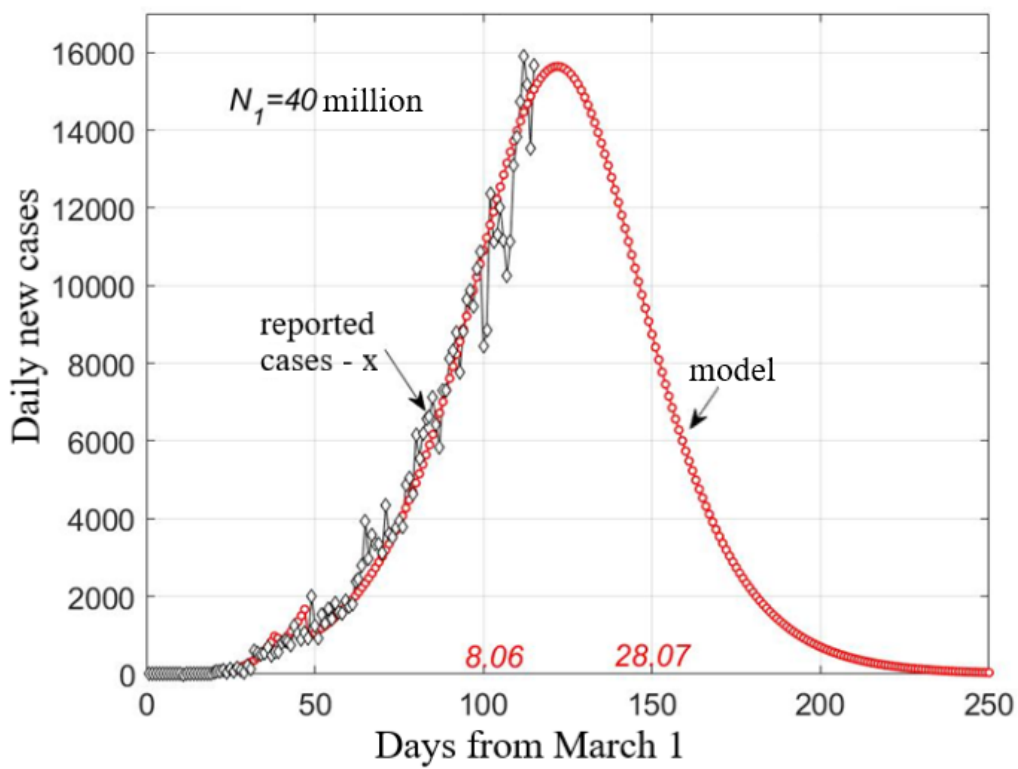

Fig. 5. Coronovirus COVID-19 epidemic development in India. The results of the daily increase in the number of infected. Where ◊ – actual data [11]

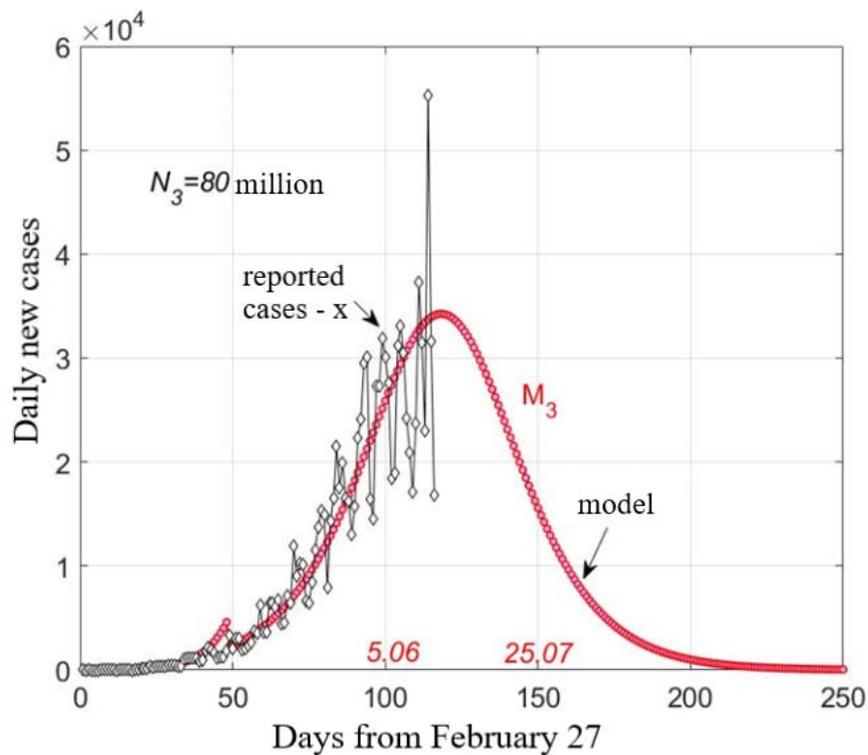

Fig. 6. COVID-19 epidemic development in Brasil. The results of the daily increase in the number of infected. Where ◊ – actual data [11]

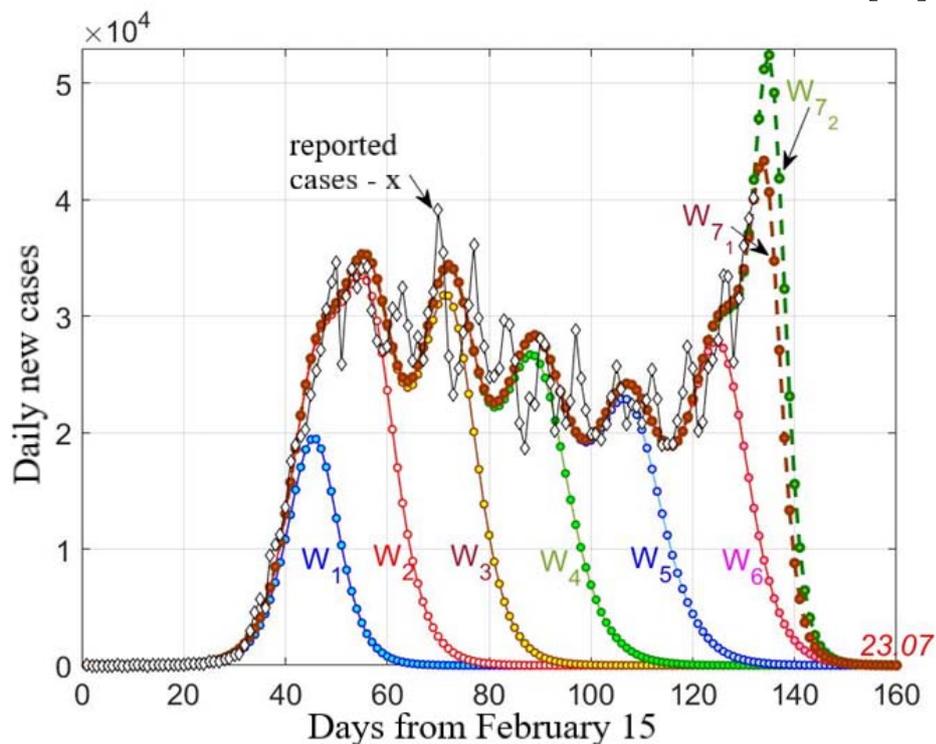

Fig. 7. All the stages of coronovirus epidemic development in the USA. The results of the daily increase in the number of infected. Where ◊ – actual data [11]; $W_1$ – the results of the 1-st wave; $W_2$ – the results of superposition of the 1-st and 2-nd waves; $W_3$ – the result of superposition of the 1, 2 and 3-d waves; $W_4$ – the results of superposition of the 1, 2, 3, 4-th waves; $W_5$ – the results of superposition of the 1, 2, 3, 4 and 5-th waves, $W_6$ – the results of the superposition of the 1, 2, 3, 4, 5

and the 6-th waves; $W_7$ – the results of the superposition of the 1, 2, 3, 4, 5, 6 and 7-th waves

Fig. 8 presents the results for the daily increase in the number of COVID-19 infected people all over the world.

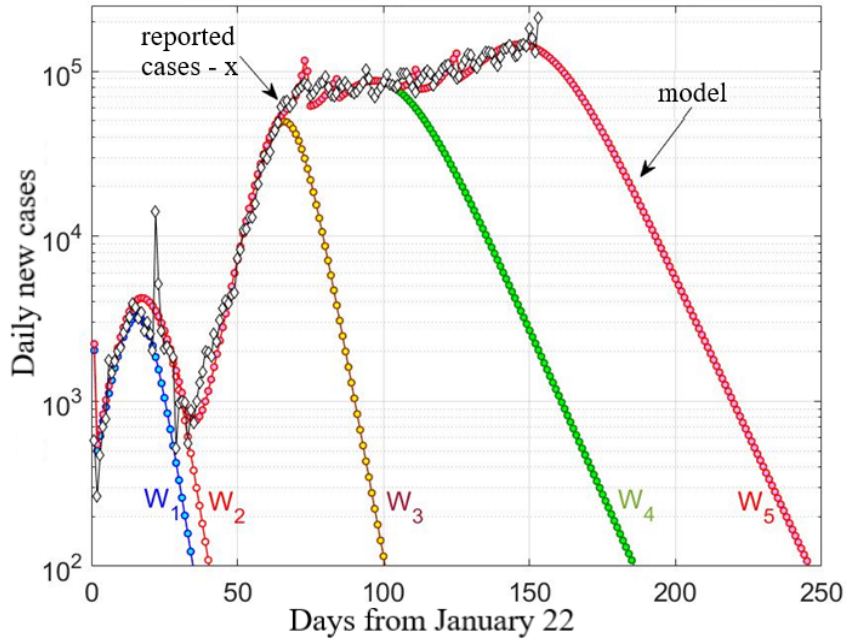

Fig. 8. All the stages of coronovirus epidemic development in the whole world. The results of the daily increase in the number of infected. Where ◊ – actual data [11]; $W_1$ – the results of the 1-st wave; $W_2$ – the results of superposition of the 1-st and 2-nd waves; $W_3$ – the result of superposition of the 1, 2 and 3-d waves; $W_4$ – the results of superposition of the 1, 2, 3, 4-th waves; $W_5$ – the results of superposition of the 1, 2, 3, 4 and 5-th waves

Table 1 shows the values for the indices of the number of infected and the capacities of the waves in Moscow.

Table 1

The values of mathematical model parameters for COVID-19 epidemic spread in Moscow

|   | 1-st wave | 2-nd wave | 3-d wave | 4-th wave | 5-th wave |
|---|---|---|---|---|---|
| $\alpha$ | 1.14 for dates: 05.03 - 22.03  1.11 for dates: 23.03 - 31.03  1.078 from 01.04 - | 1.15 from 29.03 | 1.112 from 04.04 | 1.103 from 17.04 | 1.108 from 04.05 |
| $N$ | 1,000,000 | 500,000 | 450,000 | 250,000 | 150,000 |

In Table 2 we can see the values for indices of the number of infected people and the waves capacities in Russia.

Table 2

The values of mathematical model parameters for COVID-19 epidemic spread in Russia

|   | 1-st wave | 2-nd wave | 3-d wave | 4-th wave | 5-th wave | 6-th wave |
|---|---|---|---|---|---|---|
| $\alpha$ | 1.13 for dates: 01 - 18.03<br>1.107 for dates: 19.03 – 30.03<br>1.078 for dates: 31.03 – 19.04<br>1.07 from 20.04 | 1.181 from 06.04 | 1.18 from 16.04 | 1.17 from 24.04 | 1.157 from 01.05 | 1.5 from 05.06 |
| $N$ | 3,000,000 | 500,000 | 550,000 | 550,000 | 670,000 | 1,000,000 |

Table 3 shows the values for indices of the number of infected people and the waves capacities in India.

Table 3

The values of mathematical model parameters for COVID-19 epidemic spread in India

| $\alpha$ | $N$ |
|---|---|
| 1.087 dates: 14.02 - 06.04<br>1.054 dates: 07.04 -14.04<br>1.02836 from 15.04 | 40,000,000 |

In Table 4 we can see the values for indices of the number of infected people and the waves capacities in Brasil.

Table 4

The values of mathematical model parameters for COVID-19 epidemic spread in Brasil

| $\alpha$ | $N$ |
|---|---|
| 1.161 dates: 28.02 - 19.03<br>1.067 dates; 20.03 - 15.04<br>1.0298 from 16.04 | 80,000,000 |

Table 5 presents the values for indices of the number of infected people and the waves capacities in the USA.

Table 5

The values of mathematical model parameters for COVID-19 epidemic spread in the USA

|   | 1 wave | 2 wave | 3 wave | 4 wave | 5 wave | 6 wave | 7 wave |
|---|---|---|---|---|---|---|---|
| $\alpha$ | 1.15 from 15.02 - | 1.131 from 17.02 - | 1.131 from 04.03 | 1.111 from 12.03 | 1.089 from 30.03 | 1.125 from 24.04 | 1.239 from 01.06 |
| $N$ | 2,000,000 | 4,000,000 | 4,000,000 | 4,600,000 | 4,000,000 | 3,800,000 | 1,500,000 (scenario 1) 2,000,00 (scenario 2) |

Table 6 shows the values for indices of the number of infected people and the waves capacities in the whole world.

Table 6

The values of mathematical model parameters for COVID-19 epidemic spread in the world

|   | 1 wave | 2 wave | 3 wave | 4 wave | 5 wave |
|---|---|---|---|---|---|
| $\alpha$ | 1.121 from 15.12.2019 | 1.123 from 01.01 - | 1.1048 from 17.01 | 1.2 dates: 29.02 - 05.03 | 1.2 dates: 09.04 - 11.05 |
|   |   |   |   | 1.05 dates: 06.03 - 16.03 | 1.068 dates: 12.05 - 24.05 |
|   |   |   |   | 1.045 from 17.03 - | 1.043 from 25.05 - |
| $N$ | 500,000 | 250,000 | 10,000,000 | 90,000,000 | 160,000,000 |

**4. Discussion of the results**

Figure 1 shows the calculation results on the daily increase of COVID-19 infected people, which we presented in the paper [9]. The system capacity was ~ 750,000 people. The number of COVID-19 infected people by the end of the epidemic (March) was ~ 82,000 inhabitants. In the work [9] the indices values for the number of infected population growth are presented, which decreased in accordance with the restrictive measures adopted in China. The entire population of China strictly complied with the restrictive measures prescribed by the Chinese government. For countries such as China in which restrictive measures for the epidemic spread were strictly observed, epidemic spread fits in one wave, which follows from Fig. 1 that corresponds to the daily increase in the number of infected with COVID-19 people.

Figure 2 shows the calculation results for the daily increase in number of COVID-19 infected people in Moscow, which we presented in the paper [8]. When we were modeling the epidemic spread it was at the initial stage. And having found the indices of the number of infected people growth, we did not know the value of the system capacity and considered 4 scenarios for epidemic development in Moscow. We considered the capacities: 300,000, 500,000, 700,000, and 1,000,000 Moscow residents. The capacity was to be determined in the vicinity of the peak. The scenario with the capacity of 1,000,000 people we then called New York's and by the end of the epidemic (June) it gave 75,000 infected Moscow residents. At that moment this figure for one city seemed to be awesome.

Figure 3 presents data on the daily increase in the number of people infected with COVID-19 in Moscow from the epidemic beginning to the end of June. From fig. 3 we can see that the first wave is that wave which is shown in Fig. 2 and corresponds to the scenario with a capacity of 1,000,000 people (called New York's by us). It was that wave that was the longest one, the end of this wave - the end of June and it gave us the number of infected ~ 75,000 inhabitants. The second and third waves of the epidemic spread in Moscow arose because of non-compliance with restrictive measures in the twentieth of April and led to the fact that the peak of the epidemic was shifted to May 7 and gave us another 102 thousand infected residents. The 4th wave appeared from 11.05 to 16.05 and gave us the number of infected ~ 30 thousand people. It arose due to violation of restrictions on the May holidays. The fifth wave appeared on May 26, bringing the number of coronavirus infected ~ 15 thousand people and shifting the end of the epidemic in Moscow to mid-July. The fifth wave is likely to occur when the restrictions were lifted at the beginning of industrial enterprises work in May. Thus the number of people infected in Moscow while maintaining the restrictions should have been equal to 222 thousand inhabitants at the end of the epidemic.

But since a number of restrictions were lifted in early June we examined three scenarios for the COVID-19 epidemic development when the restrictions were removed on June 9. It launched the 6th wave with the same index of the number of infected growth as for the 5th wave.

We examined 3 scenarios with different capacities: 150,000 inhabitants (as for the 5th wave), 300,000 inhabitants and 500,000 inhabitants. The peak of the 6th wave was in the vicinity of July 9, and the ending is expected in mid-August. The sixth wave will add to the number of infected people according to the scenarios, respectively: 15,000, 30,000, 50,000 inhabitants, and will bring the total number of infected to the values, respectively: 237,000, 252,000, 272,000 Moscow residents. Which of scenarios to be implemented in Moscow will be clear in mid-July.

It should be noted that the indices of the number of infected people growth and waves capacity were determined from actual data [12], the largest error in the number of infected people did not exceed 10% (relative error) and in most cases it was less then l%.

Figure 4 shows the epidemic spread in Russia (calculations dated June 25), which fits on the day of calculation for 6 waves. Data on the indices of the number of infected people growth and waves capacities of the epidemic in Russia are presented in table 2.

The local peaks of these six waves were, respectively, the dates: 29.04 (first wave), 07.05 (second wave), 20.05 (third wave), 29.05 (fourth wave), 08.06 (fifth wave), 22.06 (sixth wave). Table 7 shows the number of COVID-19 infected people that have brought these 6 waves.

Table 7

The values for the number of infected people during six waves in Russia

| 1 wave | 2 wave | 3 wave | 4 wave | 5 wave | 6 wave |
|---|---|---|---|---|---|
| 196,261 | 76,629 | 76,271 | 79,914 | 90,916 | 141,365 |

The first three waves in Russia were determined by the trend of Moscow and the Moscow region, the next three waves of COVID-19 spread in Russia are associated with the epidemic spread in the European part of Russia, the regions of the Urals, the regions of the Urals, Siberia, the Far East, and the North. Thus on June 25 the forecast according to the model of 6 waves superposition will give the number of coronavirus infected in Russia ~ 662 thousand inhabitants. Recalculation according to the mathematical model should be carried out after 10-15 days. The following waves might appear that will increase the number of infected people. It should be noted that if strict restrictive measures were observed in Russia then the country would have passed through an epidemic limited to a single wave and would have the number of infected ~ 200 thousand inhabitants as it was predicted in our work [9]. It should be mentioned that in the calculations according to the number of infected in Russia the error did not exceed 5%, in most cases it was less than 0.5 %.

Figure 5 presents a curve for the daily increase in the number of infected population in India and Table 3 shows the values of indices for the number of infected people growth and the system capacity. We can see that with the introduction of restrictions in the country, the index of infected people growth was consistently decreasing. But such a huge capacity indicates about very strong mixing in India, apparently such high capacity value is likely to be associated with large population crowding.

The system capacity is not directly related to the country population. So, the wave capacity for China is only 750 thousand people, for Russia ~ 6.3 million people (which is also not rather small). The wave peak of the epidemic spread in India is expected on June 30 with the increase of infected people not less than 15 thousand inhabitants.

The epidemic beginning in India was on February 14, the end of the epidemic will be in late November - early December. The size of a single wave will give 1,103,000 infected India residents. But we believe that in the future there may be other waves that can increase the number of infected people. The error in calculating the number of infected in India did not exceed 25% in some cases.

Figure 6 presents data on the daily increase in the number of infected COVID-19 in Brazil (calculated on June 22) Table 4 shows the data on indices of the number of infected people growth in Brazil and the system capacity. Capacity for Brazil is about 80 million inhabitants. At the moment it is the largest capacity for a single wave. Such a high capacity provides the single wave length from February 28 to the beginning of December. The peak of the wave is June 24, with the number of infected people being more than 35 thousand inhabitants. The first wave gives the number of infected people by the epidemic end about 2,307,000 Brazilians. We believe that there will be the next waves which can significantly increase the number of infected in Brazil. Although from the table 4 swe can see that the index of the number of infected people decreased during the course of the epidemic, but the measures taken in the country are apparently insufficient to curb the epidemic in the country. Capacity of 80 million people indicates about a very strong mixing in the country. The error in calculating the number of infected in some cases does not exceed 25%.

Figure 7 and Table 5 shows the calculated curves for the COVID-19 epidemic spread in the USA as well as the indices for the number of infected people growth and the waves capacity for the epidemic spread in the USA. The COVID-19 epidemic development in the United States fits into the superposition of 7 waves with total capacity of 24.4 million inhabitants. This is the third largest system capacity after the capacity of Brazil and India. This also indicates high mixing in the country and non-compliance with restrictive measures.

Table 8 shows the waves characteristics for the epidemic spread in the United.

Table 8

Waves characteristics for the epidemic spread in the United

| The wave number | Date of the wave appearance | The peak of the wave increase | | The number of infected people |
|---|---|---|---|---|
| | | date | increase | |
| 1-st wave | 02.03 | 31.03 | 19,444 | 260,869 |
| 2-nd wave | 07.03 | 10.04 | 30,309 | 463,306 |
| 3-d wave | 25.03 | 26.04 | 30,309 | 463,306 |
| 4-th wave | 03.04 | 13.05 | 25,480 | 459,584 |
| 5-th wave | 22.04 | 31.05 | 21,360 | 389,106 |
| 6-th wave | 14.05 | 19.06 | 26,048 | 422,220 |
| 7-th wave Scenario №1 Scenario №2 | 09.06 09.06 | 27.06 27.06 | 34,317 42,180 | 289,346 383,185 |

The first two waves arose one after another and appeared in early March in Washington and New York (they have different capacities) and brought a total of ~ 720 thousand infected people and then the epidemic spread was passing across states and cities (both in breadth and inland). The 6th wave seems to be related to the beginning of restrictions relaxation and brought 422,220 infected residents. But the seventh wave has formed recently - 09.06 and is connected in our opinion with the protest actions. The peak of the seventh wave is in the vicinity of June 27. Therefore we proposed 2 scenarios with different capacities for the epidemic development: 1,500,000 (scenario 1), 2,000,000 (scenario 2). Both scenarios fit well with the actual data on the number of infected with an error in some cases of less than 10%.

A global peak from the superposition of 5, 6, and 7 waves will give the increase of infected people of 43,335 inhabitants according to scenario 1 and 52,441 people in scenario 2. At the end of June and early July, it will become clear according to which scenario events are developing in the USA. According to scenario 1, the protests will give the number of infected about 289,346 people and 383,158 infected according to scenario 2. Thus on June 25 the model gives a forecast for the epidemic end to the late August with the number of 2,747,737 infected people (according to scenario 1) and with a number of 2,841,576 infected (according to scenario 2). If the following waves arise because of the restrictions lifting and high mixing then the number of infected people will increase and the epidemic duration will shift

Figure 8 and Table 6 show the calculations results (dated June 22) for modeling of the epidemic spread all over the world. The COVID-19 epidemic spread in the whole world fits into 5 waves, the characteristics of which being presented in Table 9.

Table 9

Waves characteristics in the whole world

| The wave number | Wave appearance | Date for wave peak | Increase of infected at the peak | The number of infected people | Date of the wave end |
|---|---|---|---|---|---|
| 1-st wave | 22.01 | 07.02 | 3,255 | 53,967 | 15.03 |
| 2-nd wave | 22.01 | 13.02 | 1,682 | 28,380 | 18.03 |
| 3-d wave | 10.02 | 27.03 | 49,671 | 948,586 | 21.05 |
| 4-th wave | 14.03 | 27.04 | 87,189 | 3,875,593 | 12.09 |
| 5-th wave | 21.04 | 16.06 | 146,170 | 6,802,485 | 15.11 |

The first two waves of the COVID-19 epidemic spread in the world correspond to the epidemic spread in China with the number of infected being 82,347 residents and the end on 18.03. All these data are consistent with the actual epidemic spread in China. The system capacity while calculating separately the epidemic spread in China [9] was 760 thousand inhabitants and the capacity of the first two waves in the world was 750 thousand inhabitants (see Table 9). The error in the calculations for the number of infected people in comparison with the actual data (for the first 2 waves) does not exceed 2%, and when calculating the remaining waves it does not exceed 6%.

The third wave occurred on 10.02 and ended in late May and brought the world a little less than a million infected people (948,586 see table 9). We believe that the third wave covers the epidemic spread in Europe. The fourth wave of the epidemic spread was determined on March 14 and gave a local peak on April 27. The end of the fourth wave is mid-September (see Table 9) and it will bring 3,875,593 infected people. The 4th wave is likely to capture the epidemic spread in the countries such as Russia, some states in the USA, etc.

The fifth wave was confirmed on April 21, its local peak occurred on June 16 with an increase of infected being 146,170 people (see Table 9), the end date is November. The fifth wave is most likely to cover the epidemic spread in Latin America (Brazil, Peru, Chile, etc.), in the countries of South Asia (India, Pakistan, Bangladesh, etc.), in some states of the USA.

The fifth wave is slightly less than 7 million infected residents, see table 9 (6,802,485 inhabitants). With the advent of the fifth wave the epidemic spread in the world has passed into a "new phase", the daily increase in the number of infected people in the world has exceeded 100,000 people daily. Thus, the mathematical model for COVID-19 epidemic in the world predicts (calculated on June 22) that the superposition of five waves will give the world 11,709,311 infected people.

We believe that the world expects the following waves which have not been observed in actual data yet, the next waves will increase the number of infected people in the world.

**Conclusions**

On the base of a system of discrete logistic equations we developed an approach by superposition of waves for modeling of the COVID-19 epidemic both in the whole world and in particular countries. In countries that strictly enforce restrictive measures and have small "capacities", the epidemic spread fits on one single wave (case study is the epidemic spread in China).

The COVID-19 epidemic spread in the whole world fits into the superposition of 5 waves (calculated on June 22). In further calculations to correct the model only the fifth wave will be corrected (the remaining waves will remain unchanged) and the subsequent waves will take their place corresponding to the epidemic spread in the world.

The first 2 waves characterize the epidemic spread in China; the third wave reflects the impact of the epidemic spread in European countries; the fourth wave mainly reflects the epidemic spread in the United States and Russia; the fifth wave is most likely to characterize the epidemic spread in Latin America and South Asia.

At the moment the mathematical model predicts the end date for the COVID-19 epidemic in the world not earlier than December with infected population being of over 12 million people. Subsequent waves will shift the deadline for pandemic ending in the world and increase the number of infected people.

The developed approach made it possible to determine that the COVID-19 epidemic spread in the United States fits on 7 waves. The seventh wave of the epidemic arose most likely because of the restrictions lifting and protests in the United States. The forecast according to the mathematical model is the deadline for epidemic ending in the United States will be not earlier than August with the number of infected people exceeding 2,800,000 inhabitants. Subsequent waves in the United States may lead to an increase in the number of infected people and can shift the epidemic end.

Mathematical modeling shows that at the moment the epidemic spread in Brazil and India fits into a single wave of large capacity. The epidemic end dates in Brazil and India are not earlier than November with the numbers of infected: more than 2,300,000 inhabitants in Brazil and more than 1 million inhabitants in India.

Subsequent waves that can increase the number of infected at times are not excluded.

It was shown that the epidemic spread in Russia fits into the superposition of 6 waves. Whereas the spread of the epidemic in Moscow and the Moscow Region influenced the epidemic spread in whole Russia in the first three waves, the subsequent waves are associated with the epidemic spread beyond the Urals, to Siberian regions, regions of the Far East and North in Russia. The end date of the epidemic in Russia is not earlier than August with infected population exceeding 660 thousand inhabitants.

Subsequent waves might lead to increase in the number of infected and to a shift in epidemic end in Russia. The parameters for mathematical models of the epidemic spread in the world and in the countries that give the largest increase in the number of infected people were obtained.

**List of literature**